\documentclass[%
  aps,
  preprint,
  amsmath,
  amssymb
]{revtex4-1}

\usepackage[T1]{fontenc}
\usepackage[utf8]{inputenc}
\usepackage{bm}
\usepackage{graphicx}
\usepackage[colorlinks=true,linkcolor=blue,citecolor=magenta,urlcolor=blue]{hyperref}

\begin{document}

\title{Strong CP as an Infrared Holonomy: The $\theta$ Vacuum and Dressing in Yang-Mills Theory}

\author{J.~Gamboa}
\affiliation{Departamento de Física, Universidad de Santiago de Chile, Santiago, Chile}
\email{jorge.gamboa@usach.cl}

\author{N.~Tapia-Arellano}
\affiliation{Department of Physics and Astronomy, Agnes Scott College, Decatur, GA. 30030, USA}
\email{narellano@agnesscott.edu}

\begin{abstract}
We reformulate the strong $CP$ problem from an infrared viewpoint in which the vacuum angle $\theta$ is not treated as a local coupling but as a global Berry-type holonomy of the infrared-dressed state space over $\mathcal{A}/\mathcal{G}$.
Infrared dressing is described as adiabatic parallel transport of physical states in configuration space, generated by an infrared connection $\mathcal{A}_{\rm IR}$.
Using the Chern-Simons collective coordinate, we show that the Pontryagin index emerges as an integer infrared winding, such that the resulting holonomy phase is quantized by $Q\in\mathbb Z$ and reproduces the standard weight $e^{i\theta Q}$.
A quantum rotor provides a controlled infrared example illustrating why broad classes of local correlators may remain insensitive to $\theta$, while global
response functions, such as the vacuum energy curvature and the topological susceptibility, retain a nontrivial dependence.
We contrast this picture with recent claims of $\theta$--independence based on the order of limits and show that it is consistent with both the rotor benchmark and the classic Witten-Veneziano perspective.
\end{abstract}

\maketitle

\section{Introduction}

In non-Abelian gauge theories, CP violation is not associated with local
processes or perturbative interactions, but is instead tied to the global structure of the quantum vacuum
\cite{Belavin1975,JackiwRebbi1976,tHooft1976,Witten1979,Veneziano:1979ec,PecceiQuinn1977}.
Already in pure Yang-Mills theory, the existence of topologically distinct sectors in configuration space allows for the introduction of a topological term proportional to the vacuum angle $\theta$, which fixes a global orientation of the vacuum and makes explicit CP violation possible in principle.
The experimental absence of strong CP effects suggests \cite{PDG2024} that the physical vacuum must lie extraordinarily close to a genuinely CP-invariant state, thereby turning the vacuum angle $\theta$ into a problem of ground-state selection rather than an ordinary dynamical parameter.
This gives the problem direct physical content: it is not merely a question of suppressing particular processes but of understanding why the state of minimal energy in a fundamental theory exhibits such a special global orientation \cite{Witten1979,Veneziano:1979ec}.

From this perspective, the various strategies developed in the literature share a central idea: any resolution must take seriously the deeply non-perturbative nature of a non-Abelian gauge theory in the infrared sector.
In other words, a viable resolution must identify a mechanism that allows the vacuum angle to relax or to be dynamically selected, either through explicit dynamical ingredients or through geometric and topological conditions that govern the implementation of CP as a global quantum symmetry
\cite{Belavin1975,tHooft1976,JackiwRebbi1976,Witten1979,Veneziano:1979ec,PecceiQuinn1977,Weinberg1978,Wilczek1978}.

Within this context, several approaches have been discussed.
More recently, a series of works has explored the possibility that CP is an exact discrete symmetry at the quantum level, whose spontaneous breaking --under appropriate global conditions-- would give rise to the observed physics
\cite{AiCruzGarbrechtTamarit2022, AiGarbrechtTamarit2024,AlbandeaCatumbaRamosPRD}.
In this work, we adopt a closely related starting point but propose a complementary interpretation based on geometric principles: the angle $\theta$ is understood as a holonomy of the infrared vacuum associated with adiabatic transport in the functional space of gauge fields \cite{Gamboa:2025qjr,Gamboa:2025dry,Gamboa:2025fcn,Gamboa:2025nco}.
This viewpoint frames CP violation as a global property of the vacuum rather than as a local dynamical effect.

The central infrared statement of this work can be summarized succinctly: \emph{infrared dressing is holonomy}.
Physical infrared states form a nontrivial bundle over $\mathcal{A}/\mathcal{G}$, and adiabatic evolution along a closed contour $C\subset\mathcal{A}/\mathcal{G}$ produces a Berry holonomy $\mathcal{U}_C$ that cannot be removed globally.
The associated phase is topological: it is controlled by an integer winding number (the Pontryagin index), so $Q\in\mathbb Z$ discretizes the holonomy and implies the standard $2\pi$ periodicity in $\theta$.

In this paper, we develop a non-perturbative infrared formulation in which the vacuum angle $\theta$ appears as a Berry holonomy of the infrared-dressed state space over $\mathcal{A}/\mathcal{G}$.
We relate this holonomy to the Chern-Simons collective coordinate and show how the Pontryagin index arises as an integer infrared winding.
This framework clarifies why $\theta$ survives as a global datum of the infrared representation under admissible deformations, even when broad classes of local observables remain insensitive to it.

\section{Yang-Mills Theory with a $\texorpdfstring{\boldsymbol{\theta}}{theta}$ term and topological sectors}

Consider pure Yang-Mills theory in four dimensions, supplemented by the
topological $\theta$ term \cite{Jackiw:1977yn},
\begin{equation}
S[A]
=
\frac{1}{2g^{2}}\int d^{4}x\,\mathrm{tr}\,F_{\mu\nu}F^{\mu\nu}
\;+\;
i\,\theta\, Q[A],
\qquad
Q[A]
:=
\frac{1}{32\pi^{2}}\int d^{4}x\,\mathrm{tr}\,F_{\mu\nu}\widetilde F^{\mu\nu},
\label{eq:YM_theta_action}
\end{equation}
where $\widetilde F^{\mu\nu}=\tfrac12\epsilon^{\mu\nu\rho\sigma}F_{\rho\sigma}$.
For gauge field configurations of finite Euclidean action, the Pontryagin density defines a topological invariant: the Pontryagin index (second Chern number),
\begin{equation}
Q[A]\in\mathbb{Z}.
\label{eq:Pontryagin_integer}
\end{equation}
Equivalently, finite-action configurations approach pure gauge at infinity, $A_\mu\to g^{-1}\partial_\mu g$ on $S^3_\infty$, so that gauge fields are classified by homotopy classes labeled by the winding number $n\in\pi_3(G)\simeq\mathbb{Z}$ (for $G=SU(N)$).

As a consequence, the Yang-Mills functional integral naturally decomposes into topological sectors, and the vacuum functional takes the form
\begin{equation}
Z(\theta)
=
\sum_{n\in\mathbb{Z}} Z_n\,e^{\,i\theta n},
\qquad
Z_n
:=
\int_{\mathcal{A}_n/\mathcal{G}}\!\!\mathcal{D}A\;
\exp\!\left[
-\frac{1}{2g^{2}}\int d^{4}x\,\mathrm{tr}\,F_{\mu\nu}F^{\mu\nu}
\right],
\label{eq:Ztheta_sector_sum}
\end{equation}
where $\mathcal{A}_n$ denotes the space of gauge connections with
$Q[A]=n$.
The $\theta$ term, therefore, does not modify the local equations of motion but assigns relative phases to nonequivalent homotopy classes, thereby encoding global information about the gauge configuration space.

The CP-violating character of the $\theta$ term is already explicit at the level of the action since the density $\mathrm{tr}\,F_{\mu\nu}\widetilde F^{\mu\nu}$ is odd under CP.
Accordingly, CP violation in non--Abelian gauge theories is not a perturbative or local effect but rather a manifestation of the global structure of the quantum vacuum.

Importantly, this formulation is still largely kinematical. While $\theta$ clearly encodes global topological information through the relative phases $e^{i\theta n}$, it does not by itself specify how this information is represented dynamically in the space of physical states.
In the infrared regime, where physical states are necessarily dressed by long-wavelength gluonic configurations and a Fock--space description breaks down, one must ask how the topological information carried by $\theta$ is implemented in an effective description of low-energy degrees of freedom.

Addressing this question requires isolating the collective infrared modes that encode motion between topological sectors. This leads naturally to an effective description in terms of a reduced set of degrees of freedom, which we now develop.

\section{Infrared Dressing and the Physical Meaning of the $\theta$ Vacuum}
\label{sec:IR_theta_dressing}

The discussion above shows that the $\theta$ angle encodes global topological information by assigning relative phases to distinct homotopy sectors. We now address the corresponding infrared question: how is this global information represented once local gauge fluctuations have been integrated out and physical states are dominated by collective, long-wavelength configurations?

In a non-Abelian gauge theory, the infrared sector contains a collective mode associated with motion between topological sectors.
This mode can be isolated by integrating out non-topological degrees of freedom, leading to an effective description in terms of a single compact coordinate $\phi\sim\phi+2\pi$, which parametrizes the topological collective degree of freedom. The resulting infrared dynamics is governed by an effective Hamiltonian of the form
\begin{equation}
H_{\rm eff}
=
\frac{1}{2}
\Bigl(
 -i\partial_\phi - \frac{\theta}{2\pi}
\Bigr)^{2}
+
V(\phi),
\label{eq:rotor_effective}
\end{equation}
where $V(\phi)$ is a periodic potential reflecting the compactness of the reduced configuration space. When non-perturbative effects generate a potential of the form $V(\phi)\propto 1-\cos\phi$, the reduced system is precisely equivalent to a quantum rotor (or quantum pendulum).

In this effective description, $\theta$ appears as a background connection that shifts the canonical momentum conjugate to $\phi$.
Although it does not affect the local equations of motion, it fixes the global boundary conditions of the wavefunction on the compact configuration space, thereby determining the physical spectrum.
The familiar expression of the $\theta$ vacuum as a superposition of topological sectors,
\begin{equation}
|\,\Omega_\theta\rangle
=
\sum_{n\in\mathbb{Z}} e^{i\theta n}\,|\,\Omega_n\rangle ,
\label{eq:theta_vacuum_standard}
\end{equation}
is thus understood as the manifestation of this underlying holonomy in the infrared effective theory.

This formulation is essential because it transcends a simple interpretation based on bare or Fock vacuum states.
The very existence of a non-trivial $\theta$ angle signals that physical states in the infrared cannot be described within a Fock-space representation. Instead, both the vacuum and all physical excitations are intrinsically infrared-dressed \cite{chung,kibble1,kibble2,kibble3,kibble4,KF}  by long-wavelength gluonic configurations.
The phases $e^{i\theta n}$ are physically meaningful precisely because they are encoded in dressed states and cannot be eliminated by local field redefinitions.

From this perspective, the strong CP problem is most naturally formulated as a question about the structure of the infrared vacuum and the manner in which CP-violating phases are implemented in the space of physical, infrared-dressed states, rather than as a problem of local dynamics or perturbative corrections.

Finally, we stress that the Pontryagin term itself should not be viewed as a dynamical effect generated in the infrared.
Rather, it is a structural consequence of gauge invariance and non--linearity: in four dimensions, locality and gauge invariance allow both the Yang-Mills kinetic term and the topological density
$\mathrm{tr}\,F_{\mu\nu}\widetilde F^{\mu\nu}$.
The significance of this becomes clear only when the theory's infrared structure is taken seriously.
This stands in sharp contrast to Abelian gauge theories, where linearity and trivial topology imply a well-defined Fock description in the infrared and the absence of a non-trivial vacuum structure.

\subsection{Infrared Dressing, Holonomy, and the Pontryagin Term}
\label{subsec:IR_dressing_holonomy}

Building on the infrared effective description above, we now reformulate the dressing of physical states in geometric terms.
Our goal is to clarify how the infrared collective mode and the associated $\theta$ dependence emerge from the global structure of the space of gauge configurations, without introducing additional dynamics.

In a non-Abelian gauge theory, the intrinsic non-linearity of the field equations implies that low-energy states necessarily involve long-wavelength gluonic configurations.
These infrared gluon clouds are not perturbative corrections; rather, they define the representation of the algebra of observables in which the theory is realized.
Physical states are therefore most naturally described as sections of a nontrivial bundle over the space of gauge connections modulo gauge transformations $\mathcal{A}/\mathcal{G}$.

Operationally, an infrared dressing is encoded as parallel transport in configuration space.
Given a slow infrared contour $C\subset\mathcal{A}/\mathcal{G}$, a bare state  $|\Psi\rangle_{\rm bare}$ is mapped to a physical state according to \cite{Gamboa:2025qjr,Gamboa:2025dry,Gamboa:2025fcn,Gamboa:2025nco}
\begin{equation}
|\Psi\rangle_{\rm phys}(C)
\;=\;
\mathcal{U}_C\,|\Psi\rangle_{\rm bare},
\qquad
\mathcal{U}_C
=
\mathcal{P}\exp\!\left(
 i\oint_C \mathcal{A}_{\rm IR}
\right),
\label{eq:IR_dressing_holonomy}
\end{equation}
where $\mathcal{A}_{\rm IR}$ denotes the infrared adiabatic (Berry) connection induced on the bundle of physical states. By construction, $\mathcal{U}_C$ is a holonomy in the space of infrared--dressed states over $\mathcal{A}/\mathcal{G}$.
In this sense, infrared dressing is not an additional ingredient: it \emph{is} the Berry holonomy associated with adiabatic transport in configuration space. The crucial point is that this bundle is not globally trivial: there exist closed contours $C$ for which $\mathcal{U}_C$ cannot be removed by any local redefinition of states.
Such holonomies label inequivalent infrared representations and encode physically meaningful global phases.

The quantization of the infrared Berry flux follows directly from topology \cite{Berry1984,WilczekZee1984}. Closed contours $C\subset\mathcal A/\mathcal G$ fall into homotopy classes classified by $\pi_1(\mathcal A/\mathcal G)\simeq\pi_3(G)=\mathbb Z$.
As a consequence, the Berry curvature $\mathcal F_{\rm IR}=d\mathcal A_{\rm IR}$ has a quantized flux,
\begin{equation}
\frac{1}{2\pi}\int_{\Sigma}\mathcal F_{\rm IR}\;\in\;\mathbb Z,
\label{eq:IR_flux_quantization}
\end{equation}
and the holonomy along any closed contour depends only on the associated winding number $Q$,
\begin{equation}
\mathcal U_C=\exp(i\,\theta\,Q).
\label{eq:IR_holonomy_quantized}
\end{equation}
This quantization is purely topological and does not rely on any semiclassical or instanton approximation.

Within this geometric framework, the emergence of the Pontryagin term follows in an essentially kinematical way.
In four dimensions, one may introduce the Chern--Simons current $K^\mu$, defined (up to normalization conventions) through the identity
\begin{equation}
\partial_\mu K^\mu
=
\frac{1}{32\pi^2}\,
\mathrm{tr}\,F_{\mu\nu}\widetilde F^{\mu\nu},
\qquad
\widetilde F^{\mu\nu}
=
\tfrac12\epsilon^{\mu\nu\rho\sigma}F_{\rho\sigma}.
\label{eq:dK_equals_FFtilde}
\end{equation}
For gauge fields of finite Euclidean action, configurations approach pure gauge at $\tau\to\pm\infty$, so that asymptotic vacua are characterized by an integer Chern-Simons number,
\begin{equation}
N_{\rm CS}[A]
:=
\int d^3x\,K^0 .
\label{eq:NCS_def}
\end{equation}
Integrating Eq.~\eqref{eq:dK_equals_FFtilde} over Euclidean spacetime yields the Pontryagin index,
\begin{equation}
Q[A]
=
\frac{1}{32\pi^2}\int d^4x\,
\mathrm{tr}\,F_{\mu\nu}\widetilde F^{\mu\nu}
=
N_{\rm CS}(+\infty)-N_{\rm CS}(-\infty)
\;\in\;\mathbb{Z},
\label{eq:Q_as_DeltaNCS}
\end{equation}
which measures the net infrared winding between topologically inequivalent vacua.

From the infrared point of view, it is natural to isolate a compact collective coordinate $\phi\sim\phi+2\pi$ that parametrizes this winding.
Its integer winding number reproduces the Pontryagin index,
\begin{equation}
n
=
\frac{1}{2\pi}\oint_C d\phi
\;\in\;\mathbb{Z},
\qquad
n \equiv Q .
\label{eq:phi_winding_equals_Q}
\end{equation}
Because $\phi$ is compact, the only gauge-invariant information that can be associated with it is a holonomy. Accordingly, $\theta$ is implemented as a background connection along the $\phi$ direction,
\begin{equation}
\mathcal{A}_\phi = -\frac{\theta}{2\pi},
\label{eq:Aphi_theta}
\end{equation}
so that a loop winding $n$ times acquires the phase
\begin{equation}
\exp\!\left(
 i\oint_C d\phi\,\mathcal{A}_\phi
\right)
=
\exp\!\left(i\theta n\right)
=
\exp\!\left(i\theta Q\right).
\label{eq:theta_holonomy_phi}
\end{equation}

The quantization enters sharply at this point. Because the winding (Pontryagin) number is an integer, $Q\in\mathbb{Z}$, the infrared Berry phase accumulated along a closed contour is discretized by this
topological charge:
\begin{equation}
\gamma_C(\theta)
=
\arg\,\mathcal{U}_C
=
\oint_C d\phi\,\mathcal{A}_\phi
=
-\theta\,Q
\quad(\mathrm{mod}\ 2\pi).
\label{eq:Berry_phase_quantized_by_Q}
\end{equation}
Equivalently, the Berry \emph{flux} through a surface spanning $C$ is
topological and quantized, reflecting the fact that the relevant homotopy class is labeled by an integer.
This is the precise sense in which the infrared dressing phase is quantized: the holonomy is controlled by the discrete topological sector $Q$, while $\theta$ specifies the global vacuum orientation.

In this formulation, $\theta$ does not act as a parameter deforming an
underlying Fock vacuum. Instead, it corresponds to the holonomy associated with infrared dressing, projected onto the unique topological collective coordinate allowed by gauge invariance and non-Abelian dynamics.
The Pontryagin term thus acquires a clear physical interpretation: it encodes the global phase structure of infrared-dressed states and labels nonequivalent infrared representations of the Yang-Mills vacuum.

A direct physical consequence follows. The vacuum angle $\theta$ does not act as a local coupling that modifies perturbative dynamics, but instead distinguishes inequivalent infrared representations characterized by their global holonomy.
As a result, only observables sensitive to the global structure of the infrared state space can probe $\theta$, while purely local or perturbative correlators are blind to it.

From this viewpoint, CP violation in non-Abelian gauge theories is not expected to manifest itself as a local effect, but rather as a property of the infrared vacuum structure.
The strong CP problem is thus reformulated as a problem of vacuum selection: which infrared-dressed representation is dynamically realized as the theory's actual ground state, and by what mechanism is its associated holonomy fixed?

\section{A Minimal Infrared Example}
\label{subsec:IR_rotor_example}

We now illustrate the infrared holonomy mechanism in a minimal and fully controlled setting.
Our purpose is not to introduce new dynamics, but to make explicit, in the simplest possible example, how a global infrared datum affects spectral properties while remaining invisible to local probes.

We consider the effective infrared Hamiltonian~\eqref{eq:rotor_effective} acting
on a compact collective coordinate $\phi\sim\phi+2\pi$.
The vacuum angle $\theta$ enters exclusively through the global structure of the state space and may equivalently be implemented as a twisted boundary condition on the wavefunction,
\begin{equation}
\psi(\phi+2\pi)=e^{\,i\theta}\,\psi(\phi),
\label{eq:twisted_bc_rotor}
\end{equation}
which makes its holonomy character manifest. In this benchmark, the quantization is the statement $n\in\mathbb{Z}$ for the winding sectors, mirroring $Q\in\mathbb{Z}$ in Yang-Mills.

For the free rotor, $V(\phi)=0$, the energy spectrum is
\begin{equation}
E_n(\theta)
=
\frac{1}{2}
\Bigl(
 n-\frac{\theta}{2\pi}
\Bigr)^2,
\qquad
n\in\mathbb{Z},
\label{eq:rotor_spectrum}
\end{equation}
showing explicitly that the ground-state energy depends smoothly on the
holonomy parameter $\theta$.
The associated topological susceptibility,
\begin{equation}
\chi_{\rm top}
=
\left.
\frac{\partial^2 E_0(\theta)}{\partial\theta^2}
\right|_{\theta=0}
=
\frac{1}{4\pi^2 },
\label{eq:rotor_susceptibility}
\end{equation}
provides a quantitative measure of the global infrared response of the system.

At the same time, expectation values of operators that are local in $\phi$ are
insensitive to the twisted boundary condition~\eqref{eq:twisted_bc_rotor}. This demonstrates a central infrared feature: global holonomies are detected only by observables sensitive to the global structure of the state space, while purely local probes remain blind to them.

Including a periodic potential,
\begin{equation}
V(\phi)=\Lambda^4\,(1-\cos\phi),
\label{eq:pendulum_potential}
\end{equation}
turns the system into a quantum pendulum. While this deformation lifts degeneracies and induces tunneling between winding sectors, it does not modify the holonomy structure encoded by $\theta$.
The dependence of the vacuum energy on $\theta$ persists, and the spectrum organizes into bands whose structure is again controlled by the global boundary condition.

This minimal laboratory captures the infrared logic underlying the Yang-Mills $\theta$ vacuum. The Chern-Simons collective coordinate plays the role of $\phi$, the Pontryagin index labels winding sectors, and the $\theta$ term implements the corresponding infrared holonomy.
Apparent $\theta$-independence of local observables thus reflects a projection onto probes that are blind to holonomy data, rather than the absence of a nontrivial infrared vacuum structure.

\section{Quantitative Benchmark: $\theta$-Dependence, Order of Limits, and the Role of Quarks}
\label{subsec:quant_compare_AGT}

Recent work by \cite{AiCruzGarbrechtTamarit2022,AiGarbrechtTamarit2024} argues that the conventional formulation of the strong $CP$ problem is sensitive to the order in which the infinite spacetime volume limit is taken relative to the sum over topological sectors. For a recent complementary discussion, see \cite{Khoze:2025auv}.

In particular, they claim that taking $V\to\infty$ prior to summing over $Q\in\mathbb Z$ causes $\theta$ to drop out of fermionic correlation functions and to become unobservable, implying $CP$ conservation in QCD; see Refs.~\cite{AiCruzGarbrechtTamarit2022,AiGarbrechtTamarit2024}.

A quantitative discussion should therefore be phrased in terms of well-defined $\theta$-dependent observables.
In (pure) Yang--Mills theory one may write
\begin{equation}
Z(\theta)=\sum_{Q\in\mathbb Z} e^{i\theta Q}\,Z_Q,
\qquad
f(\theta)=-\frac{1}{V}\ln Z(\theta),
\label{eq:Ztheta_ftheta}
\end{equation}
and define the topological susceptibility as the curvature of the vacuum energy,
\begin{equation}
\chi_{\rm top}
=
\left.\frac{\partial^2 f(\theta)}{\partial\theta^2}\right|_{\theta=0}
=
\frac{1}{V}\,\langle Q^2\rangle_{\theta=0}
=
\int d^4x\,
\langle q(x)\,q(0)\rangle_{\theta=0},
\qquad
q(x)=\frac{1}{32\pi^2}\,\mathrm{tr}\,F\widetilde F.
\label{eq:chi_top_def}
\end{equation}
These relations show explicitly that $\theta$-dependence is encoded in global response functions (derivatives of $\ln Z$), which are sensitive to the distribution of topological sectors.

In full QCD, the physically meaningful $CP$-odd parameter is not $\theta$ itself but
\begin{equation}
\bar\theta=\theta+\arg\det M_q,
\end{equation}
because anomalous axial rotations shift $\theta$ by the phase of the quark mass matrix. Moreover, infrared dressing is no longer purely gluonic: the infrared state space includes both gluonic and fermionic (or hadronic) dressing channels. Accordingly, statements about $\theta$--independence of a restricted class of fermionic correlators do not directly constrain the $\bar\theta$--dependence of
global quantities such as the vacuum energy $f(\bar\theta)$ and the
topological susceptibility $\chi_{\rm top}$.

Recent work by Refs.~\cite{AiCruzGarbrechtTamarit2022,AiGarbrechtTamarit2024}
has emphasized that, in the infinite--volume limit taken prior to the sum over topological sectors, a broad class of fermionic correlation functions becomes independent of $\theta$.
From the present viewpoint, this result does not signal the absence of a nontrivial vacuum structure.
Rather, it reflects the fact that such correlators act as local probes that are insensitive to global holonomy data characterizing the infrared representation of the theory.
In other words, the order of limits considered in
Refs.~\cite{AiCruzGarbrechtTamarit2022,AiGarbrechtTamarit2024}
implicitly selects observables that are blind to infrared holonomy, without eliminating the holonomy itself.

A sharp benchmark illustrating this distinction is provided by the quantum rotor toy model, where $\theta$ is unambiguously realized as a holonomy through a twisted boundary condition.
In this setting, both the vacuum energy and the topological susceptibility can be reconstructed nonperturbatively from local correlation functions
\cite{AlbandeaCatumbaRamosPRD}, confirming that global response functions remain sensitive to the holonomy even when many local observables are blind to it.

This picture is entirely consistent with the classic Witten-Veneziano mechanism \cite{Witten1979,Veneziano:1979ec}.
That analysis singles out the curvature of the vacuum energy with respect to $\theta$, encoded in the topological susceptibility of pure Yang--Mills theory, as the physically relevant quantity controlling the $\eta'$ mass.
Crucially, the Witten-Veneziano argument does not require generic local
correlators to be $\theta$-sensitive; it only requires a nonvanishing global response of the vacuum.
In the infrared holonomy framework advocated here, $\theta$ (and in QCD, $\bar\theta$) labels inequivalent infrared-dressed representations, while $\chi_{\rm top}$ precisely measures the response of the vacuum energy to variations of this global holonomy.

\section{$\theta$ as the variable conjugate to topological charge}
\label{subsec:theta_conjugate_charge}

A useful way to sharpen the holonomy interpretation is to regard $\theta$ as the
variable canonically conjugate to the integer topological charge $Q\in\mathbb Z$.
This is already implicit in the sector decomposition
\begin{equation}
Z(\theta)=\sum_{Q\in\mathbb Z} e^{i\theta Q}\,Z_Q,
\label{eq:Ztheta_sumQ_conjugate}
\end{equation}
which is mathematically identical to a Fourier transform between an ``angle'' $\theta$ and an integer ``momentum'' $Q$.  In this sense, $\theta$ plays the role of an angular variable specifying a representation of the topological sector, rather than that of an ordinary local coupling.  The infrared collective coordinate $\phi\sim \phi+2\pi$ introduced in Sec.~\ref{sec:IR_theta_dressing} makes this conjugate structure explicit: the winding number of $\phi$ is an integer and reproduces $Q$, while $\theta$ appears as a background connection along the $\phi$ direction, shifting the momentum conjugate to $\phi$ as in Eq.~\eqref{eq:rotor_effective}.

In the adiabatic approximation --which we employ as a controlled framework to isolate the infrared sector-- slow collective infrared motion is separated from fast non-collective fluctuations, and the latter are integrated out.  The net effect is an infrared effective dynamics on $\mathcal A/\mathcal G$ in which functional momenta acquire a geometric minimal coupling to an induced Berry connection $\mathcal A_{\rm IR}$,
\begin{equation}
-i\frac{\delta}{\delta A_i^a(x)}
\;\longrightarrow\;
-i\frac{\delta}{\delta A_i^a(x)}-\big(\mathcal A_{\rm IR}\big)_i^a(x),
\label{eq:covariant_momentum_shift}
\end{equation}
so that the commutator of covariant functional momenta is governed by the associated curvature.

At the technical level, this construction relies on the familiar adiabatic suppression of transitions out of the chosen fast-sector subspace,
\begin{equation}
\langle n'|H|n\rangle \approx 0
\qquad (n'\neq n),
\label{eq:adiabatic_suppression_levels}
\end{equation}
rather than on an exact decoupling.  However, the geometric infrared framework developed here goes beyond a purely Born--Oppenheimer interpretation. Once the induced connection is re-exponentiated into a holonomy, and physical states are reorganized as infrared-dressed sections of a nontrivial bundle over
$\mathcal A/\mathcal G$, the approximate absence of mixing acquires a structural meaning. In the infrared-dressed Hilbert space, local gauge-invariant dynamics act within each topological sector, whereas the global topology of configuration space obstructs transitions between nonequivalent sectors.

In this sense, the condition \eqref{eq:adiabatic_suppression_levels} should not be viewed merely as an adiabatic assumption but as the infrared manifestation of a geometric and topological superselection structure.  The Berry connection appearing in \eqref{eq:covariant_momentum_shift} is precisely the object whose
holonomy defines the infrared dressing operator $\mathcal U_C$ in
Eq.~\eqref{eq:IR_dressing_holonomy}.

From this perspective, the results of
Refs.~\cite{AiCruzGarbrechtTamarit2022,AiGarbrechtTamarit2024}
and the Witten-Veneziano mechanism \cite{Witten1979,Veneziano:1979ec}  address complementary aspects of the same infrared structure.
The former identifies probe classes insensitive to holonomy data, whereas the latter selects a global observable that necessarily detects it. The infrared holonomy framework provides a unified geometric interpretation of both viewpoints and clarifies why the $\theta$ vacuum can remain physically meaningful even when many local correlators appear $\theta$-independent.

\section{Discussion of the Results}

The geometric picture advocated here goes one step further.  Once the connection is re-exponentiated into a holonomy, physical states are naturally organized as infrared-dressed sections of a (generically nontrivial) bundle over $\mathcal A/\mathcal G$.  In that reorganized infrared Hilbert space,
Eq.~\eqref{eq:adiabatic_suppression_levels} \emph{can be reinterpreted as an emergent structural statement}: the effective infrared dynamics acts within the physically accessible dressed sector, while transitions that would probe different topological sectors are either dynamically inaccessible or exponentially suppressed in the infrared.  In particular, $\theta$ becomes
physically measurable only through observables that probe global holonomy data (such as global response functions), whereas broad classes of local correlators, especially those insensitive to infrared boundary conditions, may remain blind to it, as illustrated by the rotor benchmark in Sec.~\ref{subsec:IR_rotor_example}.

This conjugate-variable interpretation also makes clear why $\theta$ should not be identified with a quantized topological charge itself: the quantized object is $Q$. At the same time, $\theta$ labels how different $Q$ sectors are coherently glued together in the infrared representation via the phase $e^{i\theta Q}$.

\medskip
From this perspective, the strong $CP$ problem is naturally reformulated as a problem of infrared representation and vacuum selection.  Rather than asking why a local parameter must be extremely small, one is led to ask which infrared-dressed representation, characterized by a global holonomy, is realized as the theory's actual ground state.  The novelty of the present approach lies in identifying the near-diagonality of the infrared effective dynamics in topological sectors as a geometric consequence of infrared dressing and holonomy, rather than as a purely dynamical assumption or an ad hoc axionic mechanism.

\acknowledgments
\noindent
 This research was supported by DICYT (USACH), grant number 042531GR\_REG.
The work of N.T.A is supported by Agnes Scott College.

\end{document}